\documentclass[11pt,a4paper]{article}
\usepackage{graphicx}
\begin{document}

\begin{center}
{\Large ON THE MASS FORMULA AND WIGNER AND CURVATURE ENERGY TERMS}

\bigskip \bigskip

\small

G. ROYER

\bigskip

\footnotesize

{\em Laboratoire Subatech, UMR : IN2P3/CNRS-Universit\'e-Ecole des Mines, \\
 4 rue A. Kastler, BP 20722, 44307 Nantes Cedex 03, France, \\
E-mail: royer@subatech.in2p3.fr}

\bigskip

\small

(Received \today ) 

\end{center}

\bigskip

\footnotesize

{\em Abstract.\/} The efficiency of different mass formulas derived from the liquid drop model 
including or not the curvature energy, the Wigner term and different powers
 of the relative neutron excess $I$ has 
been determined by a least square fitting procedure to the experimental atomic 
masses assuming a constant R$_{0,charge}$/A$^{1/3}$ ratio. The Wigner term and the curvature energy 
can be used independently to improve the accuracy of the mass formula. The different fits lead to a
 surface energy coefficient of around 17-18~MeV, a relative sharp charge radius r$_0$ of 1.22-1.23~fm and a 
proton form-factor correction to the Coulomb energy of around 0.9 MeV.
\bigskip


\normalsize

\section{INTRODUCTION}

The binding energies of exotic nuclei close to the proton and neutron drip lines or in the superheavy element 
region are still poorly known and the different predictions do not agree completely. Therefore continuous efforts are
 still needed to determine the nuclear masses. Within a charged liquid drop approach, 
semi-macroscopic models including a pairing energy have been firstly advanced to reproduce the experimental nuclear masses 
\cite{wei35,bet36}. The coefficients of the Bethe-Weizs\"acker mass formula have been determined once again recently
 \cite{bas05}.   
To reproduce the irregularity of the masses as functions of $A$ and $Z$ partly due to shell closings and 
proton and neutron number parity, 
macroscopic-microscopic approaches have been elaborated, mainly the finite-range liquid drop model 
and the finite-range droplet model \cite{moll95}.  
The Thomas-Fermi statistical model with a well-chosen effective interaction \cite{ms94rep,ms96} has also allowed
to reproduced accurately the nuclear masses. Microscopic 
Hartree-Fock self-consistent theories using mean-fields and Skyrme or Gogny forces and pairing correlations  
 \cite{samy02,sto05} as well as relativistic mean field calculations \cite{bend01} have also been developed to describe 
 these nuclear masses. Finally, nuclear mass systematics using neural networks have been undertaken recently \cite{ath04}.

The evolution of the nuclear binding energy with deformation and rotation governs the fission, fusion, cluster 
and $\alpha$ decay potential barriers and the characteristics of the large deformed rotating states.
 One or two-body shape sequences have to be selected to simulate the exit or entrance channels
 \cite{hasse} in the macroscopic-microscopic models. Within a generalized liquid drop model and a quasi-molecular 
shape sequence the main features of these barriers have been reproduced using,  
firstly, four basic macroscopic terms : the volume, surface, Coulomb and nuclear proximity energy contributions and, secondly,
shell and pairing energy terms to explain structure effects and 
improve quantitatively the results \cite{rr84,rr85,roy00,rm01,rg03,rbon06}.

The purpose of the present work is to determine the efficiency of different combinations of terms 
of the liquid drop model by a least square fitting procedure to the experimentally available atomic 
masses \cite{aud03} and to  study, particularly, the separated influence of the Wigner term, 
 the curvature energy and different powers of the relative neutron excess $I$ to improve 
the GLDM.  

\section{NUCLEAR BINDING ENERGY}

The nuclear binding energy B$_{nucl}$(A,Z) which is the energy needed to separate all 
the nucleons forming a nucleus is linked to the nuclear mass M$_{n.m}$ by 
\begin{equation}                 
B_{nucl}(A,Z)=Zm_P+Nm_N-M_{n.m}(A,Z).
\end{equation}
This quantity may be connected to the experimental atomic masses given in \cite{aud03} since :
\begin{equation}                 
M_{n.m}(A,Z)=M_{a.m}(A,Z)-Zm_e+B_e(Z).
\end{equation}
The binding energy B$_e$(Z) of all removed electrons is determined by \cite{lunn03}  
\begin{equation}                 
B_e(Z)=a_{el}Z^{2.39}+b_{el}Z^{5.35},
\end{equation}
with $a_{el}=1.44381\times10^{-5}$ MeV and $b_{el}=1.55468\times10^{-12}$ MeV.
 
The following expansion of the nuclear binding energy in powers of $A^{-1/3}$ and $I=(N-Z)/A$ has been considered :
\begin{eqnarray}                 
B&=&a_v \left(1-k_{v_1}\vert I \vert-k_{v_2}I^2-k_{v_3}I^4\right)A-
a_s\left(1-k_{s_1}\vert I \vert-k_{s_2}I^2-k_{s_3}I^4\right)A^{\frac {2}{3}} 
\nonumber \\
& & -a_k\left(1-k_{k_1}\vert I \vert-k_{k_2}I^2-k_{k_3}I^4\right)A^{\frac {1}{3}} 
-a_0A^0-\frac {3}{5} \frac {e^2Z^2}{r_0A^{\frac{1}{3}}}  
+f_p \frac {Z^2}{A}
\nonumber \\   
& & -W \vert I \vert+E_{pair}-E_{shell}-E_{cong}. 
\end{eqnarray}

The first term is the volume energy and corresponds to the saturated exchange force and infinite nuclear matter.
It includes the asymmetry energy term of the Bethe-Weizs\"acker mass formula via the $I^2A$ term.
 The second term is the surface energy term. It takes into account the deficit of binding energy 
of the nucleons at the nuclear surface and corresponds to semi-infinite nuclear matter. In the Bethe-Weizs\"acker 
mass formula the dependence of the surface energy on $I$ is not considered. The third term, the curvature energy,
 is a correction to the surface energy appearing when the surface energy is viewed as a result of local properties 
of the surface and consequently depends on the mean local curvature. This term is taken into account in the 
Lublin-Strasbourg drop (LSD) model \cite{lsdm}, the TF model \cite{ms96} but not in the FRLDM \cite{moll95}.
The A$^0$ term appears when the surface term of the liquid drop model is extended to include higher order terms in 
A$^{-1/3}$ and $I$.  
The fifth term is the Coulomb energy. It gives the decrease of binding energy due to the repulsion between the protons.
 In the Bethe-Weizs\"acker mass 
formula the proportionality to Z(Z-1) is prefered. For the charge radius the formula $R_{0,charge}=r_0A^{1/3}$ is assumed,  
 a more sophisticated expression has been studied in ref. \cite{roygau}.  The $Z^2/A$ term is a proton form-factor 
correction to the Coulomb energy which takes into account the finite size of the protons. The term proportional to $I$
is the Wigner energy \cite{moll95,mye77} which appears in the counting of identical pairs in a nucleus.

The pairing energy has been determined using  
\begin{equation}
\begin{array} {ccc}
E_{pair}=-a{_p}/A^{1/2}  $ for odd Z, odd N nuclei$, \\
E_{pair}=0      $  for odd A$,  \\
E_{pair}=a{_p}/A^{1/2} $  for even Z, even N nuclei$.
\end{array} 
\end{equation}
The $a{_p}=11$ value has been adopted following different fits. More sophisticated expressions exist for this pairing
energy \cite{moll95,ms96}.
The theoretical shell effects given by the Thomas-Fermi model ($7^{th}$ column of the table in \cite{ms94rep} and \cite{ms96}) 
have been retained since they 
reproduce correctly the mass decrements from fermium to $Z=112$ \cite{hof96}. They are calculated from the Strutinsky 
shell-correction method and given for the most stable nuclei in the appendix.
The sign for the shell energy term comes from the adopted definition in \cite{ms94rep}. 
It gives a contribution of $12.84~$MeV to the binding energy for $^{208}$Pb for example.
The congruence energy term is given by :
\begin{equation}                 
E_{cong}=-10MeV~exp\left(-4.2\vert I \vert\right).
\end{equation}
It represents an extra binding energy associated with the presence of congruent pairs in contrast
to the pure Wigner expression simply proportional to $I$ \cite{ms96}.

 The masses of the 2027 nuclei verifying the two following conditions have 
been used to obtain the coefficients of the different expansions by a least square fitting procedure : 
N and Z higher than 7 and the one
 standard deviation uncertainty on the mass lower than 150 keV \cite{aud03}. 
The root-mean-square deviation has been calculated using :
\begin{equation}
\sigma ^2= \frac {\Sigma \left \lbrack  M_{Th}- M_{Exp}\right 
\rbrack ^2}{n}.
\end{equation}

In  Table~\ref{tab1}, the improvement of the experimental mass reproduction when additional contributions
are added to the basic $A,~AI^2,~A^{2/3},~A^{2/3}I^2,~Z^2/A^{1/3}$ terms is clearly shown (each calculation corresponds to 
one numbered line). The curvature energy is not taken into account. The introduction of the pairing term and of the proton form factor is 
obviously needful. In contrast, the congruence energy term does not allow to lower $\sigma$ at least with the fixed 
coefficients adopted here (as in the LSD and TF models). When the coefficients before the exponential and the exponent
are free the congruence energy tends to the Wigner term since the coefficient before the exponential diminishes 
while the exponent increases. The constant term seems unnecessary. 
 The $A^{2/3}|I|$ term is useful to improve the accuracy of the expansion and is more efficient that the $A^{2/3}I^4$
 term. The Wigner term plays the major role to decrease
$\sigma$. When the Wigner term is taken into account
the introduction of the $A^{2/3}|I|,~A^{2/3}I^4$ and $A^0$ terms are ineffective. 
Thus, the very satisfactory value of $\sigma=0.60~$MeV can be reached \cite{moll95,sto05,lsdm}. 
The introduction of the Wigner term in a liquid drop model has the main drawback that it leads to an important
 discontinuity at the transition between one and two-body shapes as in fission or fusion. Indeed, when a single system 
divides into two parts the Wigner term must be evaluated separately for the two fragments and the results added. Thus 
for the same value of  $|I|$ (symmetric fission or fusion) the Wigner term will jump at scission to 2 times its
original value. The same problem exists for the Congruence energy term.
  
\begin{table*}
\caption{Coefficient values (in MeV or fm) as functions of the selected term sets 
 including or not the congruence and pairing energies and corresponding root mean square deviation.
The shell energy is taken into account.}
\begin{center}
\begin{tabular}{|c|c|c|c|c|c|} \hline
$n°$& $ a_v$& $k_{v_1}$&$k_{v_2}$ &$k_{v_3}$&   $a_s$ \\	
$1$&14.8504&-         &  1.55448 & -       & 16.1059  \\
$2$&15.7826&-         &  1.6165  & -       & 21.017  \\
$3$&15.5959&-         &  1.70507 & -      & 17.1723 \\ 
$4$&15.6184&-         &  1.70459 & -      &17.242   \\
$5$&15.5233&-         &  1.71612 & -       & 18.071  \\
$6$&15.4285&-         &  1.71066 & -       & 17.5713  \\
$7$&15.5718& -        &  1.60965 & 1.96584 & 18.0047\\ 
$8$&15.5919&-0.04582  &  1.90751 & -       & 17.8069   \\
$9$&15.447 &-         &  1.84011 & -       & 17.3581 \\
$10$&15.4647&-0.0049   &  1.8509  & -       & 17.3894 \\
$11$&15.5147&-         &  1.85055 & -       & 17.6976 \\
$12$&15.4607& -        &  1.83014 & 0.14095 & 17.3706\\ \hline
$n°$& $k_{s_1}$&$k_{s_2}$&$k_{s_3}$& $a_0$  &$W$ \\	
$1$&-        &0.93696  &-        &-       &-        \\
$2$ &-        &1.13845  &-        &-18.253 &-          \\
$3$ &-        &0.98894  &-        &-       &-        \\ 
$4$  &-       &0.98033&  -         &-       &-       \\
$5$&-        &1.40391  &-        &-       &-         \\
$6$&-        &1.39267  &-        &1.8569  &-         \\
$7$& -       &0.75244  &10.56741 &-       &-   \\ 
$8$&-0.283   &2.39131  &-        &-       &-           \\
$9$&-        &2.11347  &-        &-       &27.5488   \\
$10$&-0.03409 &2.16959  &-        &-       &25.1137   \\
$11$&-        &2.14699  &-        &-1.409  &29.0793  \\
$12$& -       &2.01948  &1.10833  &-       &26.6425   \\ \hline
$n°$& $Cong$  &$Pairing$& $r_0$& $f_p$   &$\sigma$  \\	
$1$&y         &y        & 1.2434 &2.52888&1.156  \\
$2$&y         &y        & 1.1595 &3.34645&0.936  \\
$3$&n         &n        & 1.2272 &-       &1.322 \\ 
$4$&n        &y         &1.2244 &-  &1.032\\
$5$&n         &y        & 1.2066 &1.47705&0.687  \\
$6$&n         &y        & 1.2153 &1.39388&0.684  \\
$7$&n         &y        & 1.2042 &1.24476&0.665  \\ 
$8$&n         &y        &1.2049  &1.0255 &0.628    \\
$9$&n   &y        & 1.2252 &0.95419&0.603  \\
$10$&n       &y        & 1.2234 &0.93536&0.603  \\
$11$&n         &y        & 1.2195 &0.98825&0.601  \\
$12$&n         &y        & 1.2244 &0.91559&0.602  \\ \hline
\end{tabular}
\end{center}
\label{tab1}
\end{table*}

In  Table~\ref{tab2}, the efficiency of the curvature
 energy term with different $I$ dependences is examined, disregarding the Wigner contribution. 
The introduction of only one term in $A^{1/3}$ is ineffective while the addition of  $A^{1/3}I^2$
improves slightly the results. Supplementary terms in $|I|$ to determine the volume, surface and curvature
energies allow to reach $\sigma$=0.59~MeV. They are still more efficient than $I^4$ terms.
The curvature energy term has the advantage that it is continuous at the scission point at least 
in symmetric fission. It has the disadvantage that its value (and the sign) lacks of stability.   

\begin{table*}
\caption{Coefficient values (in MeV or fm) as functions of the selected term sets 
  and corresponding root mean square deviation.
The shell and pairing energies are taken into account but not the congruence energy and 
the Wigner term.}
\begin{center}
\begin{tabular}{|c|c|c|c|c|c|}\hline
$n°$ & $a_v$& $k_{v_1}$&$k_{v_2}$ &$k_{v_3}$&   $a_s$  \\	
$1$&15.3416&-         &  1.70872 & -       & 16.8356  \\
$2$&15.3225&-         &  1.87616 & -       & 16.9627  \\
$3$&15.2954& -        &  2.39856 & -7.873 & 16.8264  \\
$4$&15.5668& 0.17993 &  1.28391 &-        & 18.5295\\ \hline
$n°$ &$k_{s_1}$&$k_{s_2}$&$k_{s_3}$&$a_k$  &$k_{k_1}$   \\	
$1$&-        &1.40978    &- &1.99142  &-        \\
$2$&-        &2.94224  &- &1.2545  &-        \\
$3$& -       &7.85795  &-73.9552&0.709782&-          \\
$4$& 1.46203 & -2.1873 &-&-2.72953 & 25.5927         \\ \hline
$n°$&$k_{k_2}$&$k_{k_3}$& $r_0$& $f_p$   &$\sigma$  \\	
$1$& -        &-        & 1.2205 &1.33865&0.68  \\
$2$&-50.7382  &-        & 1.2281 &1.28911&0.66  \\
$3$&-395.7891 &4557.60& 1.2391 &1.03047&0.61  \\
$4$&-62.900    &   -      &1.2285& 0.91998& 0.589 \\ \hline
\end{tabular}
\end{center}
\label{tab2}
\end{table*}

A good convergency of the volume a$_v$ and asymmetry volume k$_v$ constants is observed
respectively towards around 15.5~MeV and $1.8-1.9$ . The variation of the surface coefficient is larger
 but a$_s$ evolves around 17-18~MeV. Small values of the surface coefficient favors quasi-molecular or 
two-body shapes at the saddle-point of the potential barriers while large values of $a_s$ promote 
one-body elongated shapes. The value of the proton form factor correction tends to 0.92~MeV (this value has been 
retained also in the LSD model).

The reduced charge radius r$_{0, charge}$ converges to 1.22-1.23~fm. This is in full agreement with the 
set of 799 ground state nuclear charge radii presented in ref. \cite{ang04}. In this compilation a value of 
$0.9542A^{1/3}$ for the rms charge radius is obtained, which leads to $1.23A^{1/3}$ for the effective sharp charge radius. 
In this adjustment to the nuclear masses the nuclear mass radius is not fitted. Root-mean-squared matter radii are given in 
ref. \cite{lima04} for specific nuclei.

For the Bethe-Weizs\"acker formula the fitting procedure leads to  
\begin{eqnarray}                 
B_{nucl}(A,Z)=15.69A-17.6037A^{2/3}\\
-0.71660\frac {Z(Z-1)}{A^{1/3}}-23.6745I^2A 
+E_{pair}-E_{shell} \hfill \nonumber
\end{eqnarray}
with $\sigma$=1.35~MeV. That leads to r$_0$=1.2057~fm and k$_v$=1.5089. The non dependence of the surface energy term
on the relative neutron excess $I$ explains the $\sigma$ value.

\begin{figure}[bth]
\begin{center}
\includegraphics[height=5.5cm]{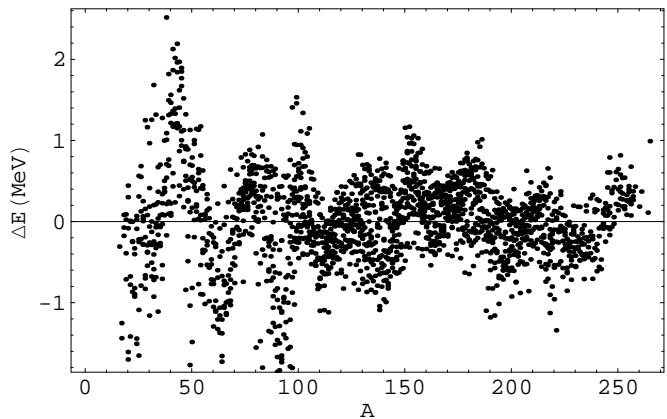}
\caption{Difference (in MeV) between the theoretical and experimental masses 
for the 2027 nuclei as a function of the mass number.}
\label{rrp}
\end{center}
\end{figure}

The Fig.~\ref{rrp} shows the dispersion between the theoretical and experimental masses within the 
last formula presented in Table~\ref{tab2} and given below : 
\begin{eqnarray}                 
B&=&15.5668\left(1-0.17993\vert I \vert-1.28391I^2\right)A
\nonumber \\
& &-18.5295\left(1-1.46203\vert I \vert
+2.1873I^2\right)A^{\frac {2}{3}}
\nonumber \\
& &+2.72953\left(1-25.5927\vert I \vert+62.9I^2\right)A^{\frac {1}{3}} 
\nonumber \\
& &-\frac {3}{5} \frac {e^2Z^2}{1.2285A^{\frac{1}{3}}}+0.91998\frac {Z^2}{A} +E_{pair}-E_{shell}.
\end{eqnarray}

\section{SUMMARY AND CONCLUSION}
The efficiency of different mass formulas derived from the liquid drop model 
and including or not a curvature energy term, the Wigner term and different powers
 of the relative neutron excess $I$ has 
been determined by a least square fitting procedure to 2027 experimental atomic 
masses assuming a constant R$_{0,charge}$/A$^{1/3}$ ratio. The Wigner term and the curvature energy term can improve 
independently the accuracy of the mass formula. The very satisfactory value of $\sigma=0.59~$MeV can be reached.
The different fits lead to a volume energy coefficient of around 15.5~MeV, a
 surface energy coefficient of around 17-18~MeV, a relative charge radius r$_0$ of 1.22-1.23~fm and a 
proton form-factor correction of around 0.9 MeV. The addition of a term in $|I|$ in the volume, surface and curvature 
energy terms is more efficient than a term in $I^4$.

\section{APPENDIX}
\begin{table*}[tbh]
\caption{Theoretical shell energy (in MeV) extracted from \cite{ms94rep} ($7^{th}$ column) at the ground state of nuclei
 for which the half-life is higher than 1 ky.}
\begin{center}
\begin{tabular}{|c|c|c|c|c|c|c|c|}\hline
$^{16}O$&$^{17}O$&$^{18}O$&$^{19}F$&$^{20}Ne$& $^{21}Ne$&$^{22}Ne$&$^{23}Na$\\	
-0.45  &1.19  &1.3  & 2.76 &  2.81  &2.82  &2.19 &  2.22      \\
$^{24}Mg$&$^{25}Mg$&$^{26}Mg$&$^{26}Al$&$^{27}Al$&$^{28}Si$&$^{29}Si$&$^{30}Si$\\	
1.64 &1.77 & 0.64 & 1.89  &  0.79 &-0.26   & -0.25 &0.22  \\
$^{31}P$ &$^{32}S$&$^{33}S$&$^{34}S$& $^{36}S$&$^{35}Cl$&$^{36}Cl$&$^{37}Cl$\\	
 0.22 &0.66  & 0.86  & 1.13  & 1.22&  1.32 & 1.48 &1.40   \\
$^{36}Ar$&$^{38}Ar$&$^{40}Ar$&$^{39}K$&$^{40}K$&$^{41}K$&$^{40}Ca$&$^{41}Ca$ \\	
1.57  &1.64  &2.45 &1.79 &2.58&2.58 &1.71  &  2.49 \\
$^{42}Ca$&$^{43}Ca$&$^{44}Ca$&$^{46}Ca$&$^{48}Ca$&$^{45}Sc$&$^{46}Ti$&$^{47}Ti$\\
2.49 & 2.16 &1.7& 0.69&-0.82&2.43 & 2.44 &1.95\\   
$^{48}Ti$&$^{49}Ti$& $^{50}Ti$&$^{50}V$&$^{51}V$&$^{50}Cr$&$^{52}Cr$&$^{53}Cr$ \\	
1.44& 0.81 &-0.05  & 0.54 & -0.31& 0.74 & -0.69  &-0.38    \\
$^{54}Cr$ &$^{53}Mn$&$^{55}Mn$&$^{54}Fe$&$^{56}Fe$&$^{57}Fe$& $^{58}Fe$&$^{60}Fe$\\
0.74& -1.11   & 0.68      &-1.54    &0.05 & 0.72   &1.22   & 2.07\\          
$^{59}Co$&$^{58}Ni$&$^{59}Ni$&$^{60}Ni$&$^{61}Ni$&$^{62}Ni$&$^{64}Ni$&$^{63}Cu$\\	
0.77   & -1.58   & -0.68  &-0.16    &0.59   &1.1    & 1.63& 1.86         \\
$^{65}Cu$&$^{64}Zn$& $^{66}Zn$&$^{67}Zn$&$^{68}Zn$&$^{70}Zn$&$^{69}Ga$&$^{71}Ga$\\	
2.33& 2.53    & 2.89    & 3.16  & 2.99    & 2.94    & 3.79    &3.71   \\ 
$^{70}Ge$&$^{72}Ge$&$^{73}Ge$&$^{74}Ge$&$^{76}Ge$&$^{75}As$& $^{74}Se$&$^{76}Se$\\	
4.13  & 4.08   & 4.19    &  3.82   & 2.53&4.08    & 4.42     &  4.08        \\
$^{77}Se$&$^{78}Se$&$^{79}Se$&$^{80}Se$&$^{82}Se$&$^{79}Br$&$^{81}Br$&$^{80}Kr$\\	
4.06    & 3.27    & 2.87    &  1.89   &   0.38  &   4.07  &   2.28  &   4.39  \\
$^{81}Kr$&$^{82}Kr$& $^{83}Kr$&$^{84}Kr$&$^{86}Kr$&$^{85}Rb$&$^{87}Rb$&$^{86}Sr$\\	
3.77 & 2.74    & 1.65 &0.96   & -0.40   & 1.13    & -0.35   &  0.79         \\
$^{87}Sr$&$^{88}Sr$&$^{89}Y$&$^{90}Zr$&$^{91}Zr$&$^{92}Zr$& $^{93}Zr$&$^{94}Zr$\\	
 0.05   & -0.97   &  -1.19 &  -1.63  & -0.47 & 0.46    &   1.53   &  2.54  \\
$^{96}Zr$&$^{92}Nb$&$^{93}Nb$&$^{94}Nb$&$^{92}Mo$&$^{93}Mo$&$^{94}Mo$&$^{95}Mo$\\	
 3.49 & -0.57  &   0.44  &  1.51   &  -2.12  &   -1.07 & -0.12   & 0.97   \\
$^{96}Mo$&$^{97}Mo$& $^{98}Mo$&$^{100}Mo$&$^{97}Tc$&$^{98}Tc$&$^{99}Tc$&$^{96}Ru$\\	
 1.79 &  2.46    &    2.98 &  3.62    & 1.26    & 2.05    & 2.64&-1.11        \\
$^{98}Ru$&$^{99}Ru$&$^{100}Ru$&$^{101}Ru$&$^{102}Ru$&$^{104}Ru$& $^{103}Rh$&$^{102}Pd$\\	
0.57    &   1.36  &  2.00    &   2.54 &   2.98&3.49     & 2.44     &  0.63   \\
$^{104}Pd$&$^{105}Pd$&$^{106}Pd$&$^{107}Pd$&$^{108}Pd$&$^{110}Pd$&$^{107}Ag$&$^{109}Ag$\\	
1.83     & 2.39    & 2.80  &  3.08    & 3.34 &  3.42   &  2.20   &  2.91         \\
$^{106}Cd$&$^{108}Cd$& $^{110}Cd$&$^{111}Cd$&$^{112}Cd$&$^{113}Cd$&$^{114}Cd$&$^{116}Cd$\\	
  0.31 &1.35     & 2.11      &  2.42    & 2.52     & 2.61     & 2.50     &   2.26 \\
$^{113}In$&$^{115}In$&$^{112}Sn$&$^{114}Sn$&$^{115}Sn$&$^{116}Sn$& $^{117}Sn$&$^{118}Sn$\\	
 1.83    & 1.97     &  0.37  &0.81    &  1.03  & 0.94     & 0.96      &  0.75       \\
$^{119}Sn$&$^{120}Sn$&$^{122}Sn$&$^{124}Sn$&$^{126}Sn$&$^{121}Sb$&$^{123}Sb$&$^{120}Te$\\	
 0.70     & 0.19     & -0.99    & -2.51    &  -4.36   &  0.76    & -0.16    & 2.08     \\ \hline
\end{tabular}
\label{tab:table3a}
\end{center}
\end{table*}

\begin{table*}[tbh]
\begin{center}
\begin{tabular}{|c|c|c|c|c|c|c|c|}\hline

$^{122}Te$&$^{123}Te$& $^{124}Te$&$^{125}Te$&$^{126}Te$&$^{128}Te$&$^{130}Te$&$^{127}I$\\	
 1.55 & 1.30      & 0.66      &  0.25    & -0.62    & -2.46    & -4.74    &0.35   \\
$^{129}I$&$^{128}Xe$&$^{129}Xe$&$^{130}Xe$&$^{131}Xe$&$^{132}Xe$& $^{134}Xe$&$^{136}Xe$\\	
 -1.29   &   1.01   & 0.48    & -0.31    &  -1.11 & -2.36   & -4.92  &-7.2    \\
$^{133}Cs$&$^{135}Cs$&$^{132}Ba$&$^{134}Ba$&$^{135}Ba$&$^{136}Ba$&$^{137}Ba$&$^{138}Ba$\\	
 -1.28  &  -3.90   &  0.93    &  -0.55   &   -1.45  &   -3.01  & -4.18   &  -5.29     \\
$^{137}La$&$^{138}La$&$^{139}La$&$^{136}Ce$&$^{138}Ce$&$^{140}Ce$&$^{142}Ce$&$^{141}Pr$\\	
 -2.22&   -3.37  &   -4.50   &  0.70    & -1.57  &-3.86    & -2.06    & -3.26        \\
$^{142}Nd$&$^{143}Nd$&$^{144}Nd$&$^{145}Nd$&$^{146}Nd$&$^{148}Nd$& $^{150}Nd$&$^{144}Sm$\\
 -2.88    &  -2.14   & -1.04    & 0.10    &  0.56 & 0.85   &  0.54     &  -2.28   \\
$^{146}Sm$&$^{147}Sm$&$^{148}Sm$&$^{149}Sm$&$^{150}Sm$&$^{152}Sm$&$^{154}Sm$&$^{151}Eu$\\
 -0.41    & 0.67     &  1.12 &1.22    &  1.31    &   0.90    & 0.38     &  1.39        \\
$^{153}Eu$&$^{150}Gd$& $^{152}Gd$&$^{154}Gd$&$^{155}Gd$&$^{156}Gd$&$^{157}Gd$&$^{158}Gd$\\	
 1.02 &   1.30    & 1.59      &  1.33    & 1.05     & 0.89     & 0.62     &  0.56   \\
$^{160}Gd$&$^{159}Tb$&$^{154}Dy$&$^{156}Dy$&$^{158}Dy$&$^{160}Dy$& $^{161}Dy$&$^{162}Dy$\\	
0.21 & 0.64   & 1.63     &   1.56   & 1.24 & 0.92     &   0.64    &    0.47 \\
$^{163}Dy$&$^{164}Dy$&$^{165}Dy$&$^{163}Ho$&$^{165}Ho$&$^{162}Er$&$^{164}Er$&$^{166}Er$\\	
0.16  &   -0.06  & -0.41    &  0.46     &  -0.12   &  1.20    &   0.70& 0.07         \\
$^{167}Er$&$^{168}Er$& $^{170}Er$&$^{169}Tm$&$^{168}Yb$&$^{170}Yb$&$^{171}Yb$&$^{172}Yb$\\	
-0.37 &  -0.54    &   -1.08    &   -0.60  &   0.32   &   -0.34   & -0.76    &  -0.94  \\
$^{173}Yb$&$^{174}Yb$&$^{176}Yb$&$^{175}Lu$&$^{176}Lu$&$^{174}Hf$& $^{176}Hf$&$^{177}Hf$\\	
-1.32  &  -1.30   &  -1.74   &  -1.23    &   -1.62 &-0.38    &   -0.90   &    -1.33  \\
$^{178}Hf$&$^{179}Hf$&$^{180}Hf$&$^{182}Hf$&$^{181}Ta$&$^{180}W$&$^{182}W$&$^{183}W$\\	
 -1.53   & -1.97    &  -1.99   & -2.16    &   -2.02  &   -1.21 &   -1.71 &  -2.00\\
$^{184}W$&$^{186}W$& $^{185}Re$&$^{187}Re$&$^{184}Os$&$^{186}Os$&$^{187}Os$&$^{188}Os$\\	
 -2.02 &  -2.38   &   -2.19  &-2.48 &  -1.61   &   -1.88  & -2.16 &  -2.08         \\
$^{189}Os$&$^{190}Os$&$^{192}Os$&$^{191}Ir$&$^{193}Ir$&$^{190}Pt$& $^{192}Pt$&$^{194}Pt$\\	
 -2.43   &   -2.47  &  -3.50   &  -2.54   &  -3.62 &  -0.97  &   -2.01   &  -3.31     \\
$^{195}Pt$&$^{196}Pt$&$^{198}Pt$&$^{197}Au$&$^{196}Hg$&$^{198}Hg$&$^{199}Hg$&$^{200}Hg$\\	
-4.04  &   -4.80  &   -6.11  &   -5.56  & -4.51    &   -5.99  &   -6.75  &    -7.52   \\
$^{201}Hg$&$^{202}Hg$& $^{204}Hg$&$^{203}Tl$&$^{205}Tl$&$^{202}Pb$&$^{204}Pb$&$^{205}Pb$\\	
-8.37& -9.11    &   -10.69  &  -9.97   &    -11.58 &   -8.22  & -10.02  & -11.00 \\
$^{206}Pb$&$^{207}Pb$&$^{208}Pb$&$^{208}Bi$&$^{209}Bi$&$^{226}Ra$& $^{229}Th$&$^{230}Th$\\	
-11.82 &   -12.68 &  -12.84  &  -11.70  &  -11.95 &  -0.30    &   -0.52   &  -0.43   \\
$^{232}Th$&$^{231}Pa$&$^{233}U$&$^{234}U$&$^{235}U$&$^{236}U$&$^{238}U$&$^{236}Np$\\	
 -0.60    &   -0.79  &  -1.27  &   -1.23 &  -1.46 &-1.30  &   -1.27  &   -1.85  \\
$^{237}Np$&$^{239}Pu$& $^{240}Pu$&$^{242}Pu$&$^{244}Pu$&$^{243}Am$&$^{245}Cm$&$^{246}Cm$\\	
  -1.74 & -2.12    &    -1.95  &   -1.99  &   -2.08  &   -2.44  &  -3.05    &   -2.96  \\ 
$^{247}Cm$&$^{248}Cm$&$^{247}Bk$& & & & & \\	
-3.17   &    -3.00 & -3.46 & & & & &      \\ \hline
\end{tabular}
\label{tab:table3b}
\end{center}
\end{table*}

\end{document}